\documentstyle[psbox]{WORKSHOP}

\begin{document}

\title{
The Catalina Real-Time Transient Survey (CRTS) \\
}

\author{
S.G. Djorgovski$^1$, A.J. Drake$^1$, A.A. Mahabal$^1$, M.J. Graham$^1$, C. Donalek$^1$, R. Williams$^1$, \\
E.C. Beshore$^2$, S.M. Larson$^2$, J. Prieto$^3$, M. Catelan$^4$, E. Christensen $^5$, R.H. McNaught$^6$ 
\\[12pt]  
$^1$  California Institute of Technology, Pasadena, CA 91125, USA\\
$^2$  Lunar and Planetary Laboratory, Univ. of Arizona, Tucson, AZ 85721, USA \\
$^3$  Observatories of the Carnegie Institution of Washington, Pasadena, CA 91101, USA \\
$^4$  Pontificia Universidad Catolica de Chile, Santiago, Chile \\
$^5$  Gemini Observatory, Casilla 603, La Serena, Chile \\
$^6$  Siding Spring Observatory, Australian National University, Canberra, Australia\\
{\it E-mail(SGD): george@astro.caltech.edu} 
}

\abst{
Catalina Real-Time Transient Survey (CRTS) is a synoptic sky survey uses data
streams from 3 wide-field telescopes in Arizona and Australia, covering the
total area of $\sim 30,000$, down to the limiting magnitudes $\sim 20 - 21$ mag
per exposure, with time baselines from 10 min to 6 years (and growing); there
are now typically $\sim 200 - 300$ exposures per pointing, and coadded images
reach deeper than 23 mag.  The basic goal of CRTS is a systematic exploration
and characterization of the faint, variable sky.  The survey has detected
$\sim 3,000$ high-amplitude transients to date, including $\sim 1,000$
supernovae, hundreds of CVs (the majority of them previously uncatalogued), and
hundreds of blazars / OVV AGN, highly variable and flare stars, etc.  CRTS has a
complete {\it open data} philosophy: all transients are published immediately
electronically, with no proprietary period at all, and all of the data (images,
light curves) will be publicly available in the near future, thus benefiting the
entire astronomical community.  CRTS is a scientific and technological
testbed and precursor for the grander synoptic sky surveys to come.
}

\kword{sky surveys --- optical transients --- supernovae --- blazars --- cataclysmic variables}

\maketitle
\thispagestyle{empty}

\section{Exploring the Time Domain Frontier}

Exploration of the time domain of the observable parameter space is now one of
the most exciting and rapidly growing areas of astrophysics, and a vibrant new
observational frontier.  A number of important astrophysical phenomena can be
discovered and studied only in the time domain, ranging from exploration of the
Solar System to cosmology and extreme relativistic sources.  There is a real and
exciting possibility of discovery of new types of objects and phenomena: 
Opening new domains of the observable parameter space often leads to new and
unexpected discoveries (Harwit 1981; Paczynski 2000; Djorgovski et al. 2001).

The field has been fueled by the advent of the new generation of digital
synoptic sky surveys, which cover the sky many times, as well as the ability to
respond rapidly to transient events using robotic telescopes.
This new growth area of astrophysics has been enabled by information technology,
continuing evolution from large panoramic digital sky surveys, to panoramic
digital cinematography of the sky, leading towards the LSST.

The relatively small synoptic sky surveys today (like CRTS) are thus both
scientific and technological precursors and testbeds for the grander surveys in
the future, such as LSST or SKA.  Processing and analyzing real-time massive
data streams exercises methods and techniques developed and needed for the
analysis of large digital sky surveys, with an added challenge of the time
coordinate; time-domain astronomy may be the ``killer app'' of the 
Virtual Observatory framework.

\section{Survey Description}

The fundamental motivation behind this project is a systematic exploration of
the time domain in astronomy, building upon, and continuing the work we have
started in the course of the earlier Palomar-Quest (PQ) survey (Djorgovski
et al. 2008).

The Catalina Real-Time Transient Survey (CRTS; http://crts.caltech.edu)
leverages existing synoptic telescopes and image data resources from the
Catalina Sky Survey (CSS) for near-Earth objects and potential planetary hazard
asteroids (NEO/PHA), conducted by the Univ. of Arizona LPL group (S. Larson, E.
Beshore, and collaborators; see http://www.lpl.arizona.edu/css/).  CSS utilizes
three wide-field telescopes: the 0.68-m Catalina Schmidt at Catalina Station,
AZ, the 0.5-m Uppsala Schmidt (Siding Spring Survey, or SSS, in collaboration
with the Australian National University) at Siding Spring Observatory, NSW,
Australia, and the Mt. Lemmon Survey (MLS), a 1.5-m reflector located on Mt.
Lemmon, AZ.  Each telescope employs a camera with a single, cooled, 4k$\times$4k
back-illuminated, unfiltered CCD.  The combined CSS+SSS+MLS data streams can
cover up to $\sim 2,000~deg^2$ per night to a limiting magnitude of $V \sim 19 
-Ð 20$ mag, plus a smaller area (up to $200~deg^2$ per night) to a limiting
magnitude of $V \sim 21.5$ mag.  All telescopes are operated for 23 nights per
lunation, centered on new moon.  Between the three telescopes, the majority of
the observable sky is covered at least once (and up to 4 times) per lunation,
depending on the time since the area was last surveyed, and proximity to the
ecliptic.  The total area coverage is $\sim 30,000~deg^2$, and it excludes the
Galactic plane within $\vert b \vert < 10^\circ Ð- 15^\circ$).  Four images of
the same field are taken, separated in time by $\sim 10$ min, for a total time
baseline of $\sim 30$ min in that sequence.  Typically 2 to 4 such sequences are
obtained per field per lunation; the cycle is generally repeated the next
lunation, marching through the RA range during the year.  The time baselines
now extend to $\sim 6$ years with up to $\sim 300$ exposures per pointing over
much of the surveyed area so far.  This represents an unprecedented coverage in
terms of the combined area, depth, and number of epochs.

CRTS taps into these data streams, detecting astrophysical transient and
variable objects outside the Solar System.  The search is performed in the
catalog domain, but we are exploring the use of image subtraction as well.
Catalogs of the sources detected in the latest images are compared to those from
median-stacked baseline images (typically some months to years old), which
typically reach $\sim 2$ mag deeper.  We identify sources that display
significant changes in brightness, or which appear where no source was
previously detected.  The contrast threshold is deliberately set high (flux
changes of at least $\sim 1$ mag and $\sim 5~\sigma$), since the most dramatic
transients/variables are likely to be the most interesting, and we cannot
even follow-up all of those.  To date, we have catalogued $\sim 3,000$ distinct
transients (many are detected repeatedly).  Lowering of the contrast
threshold could increase the transient discovery rate by an order of magnitude.
The processing pipeline is based on an earlier one developed for the Palomar-Quest 
(PQ) survey. 

A key feature of CRTS is that it is the first fully open synoptic sky survey:
all detected transients are published immediately, with no proprietary period at
all, using several Internet-based mechanisms (see, e.g., Williams et al. 2009).
This open-data approach benefits the entire astronomical community, and maximizes
the scientific returns by encouraging follow-up by other groups.  Furthermore,
we plan to put all of the available archival data on a public server, accessible
through the standard VO protocols.  This would undoubtedly enable a number of
archival studies by the entire astronomical community, in addition to the
functionality needed for our survey itself.  About $\sim 10$\% of transients for
which we obtain follow-up observations, and/or which appear interesting on the
basis of archival data, are also published as ATel and CBET notices
(a few hundred of them as of the late 2010).
A description of the survey and some early results were presented by Drake et al. (2009).

\section{A Sampling of the Scientific Results}

CRTS is now producing a steady stream of discoveries, including Supernovae, CVs
of all types, blazars and other highly variable AGN, flare stars (e.g., UV Ceti
type), high-amplitude pulsating variables (e.g., Miras), etc.  Preliminary 
astrophysical classifications, based on the CRTS and VO archival data, 
are provided within a day for all detected transients, and posted on line at
the CRTS website.  About a quarter of them cannot be classified reliably in
this manner, but we expect that this situation will improve in time.

As of the early 2011, we have detected about 1,000 confirmed or likely SNe.
In 2009 and 2010, we published more SNe than any other survey.  Some of them,
including SN 2008fz, the most luminous SN ever discovered until very recently
(Drake et al. 2010; Fig. 1), are hyper-luminous events, which may represent the new
type of pair-instability SNe (Gal-Yam et al. 2009).

We found a number of very long-lasting SN IIn events, the most
extreme of them being SN 2008iy, which took $> 400$ days to reach the peak
(Drake et al., in prep.; Fig. 1).  One interpretation of these events is that
they originated from $\eta$ Carinae type progenitors, massive stars that have
undergone a considerable pre-explosion mass loss, where the SN shock propagates
through the stellar wind ejecta for a considerable time, converting its
kinetic energy into the extended light curve.

\begin{figure*}[t]
\centering
\psbox[xsize=16.5cm]
{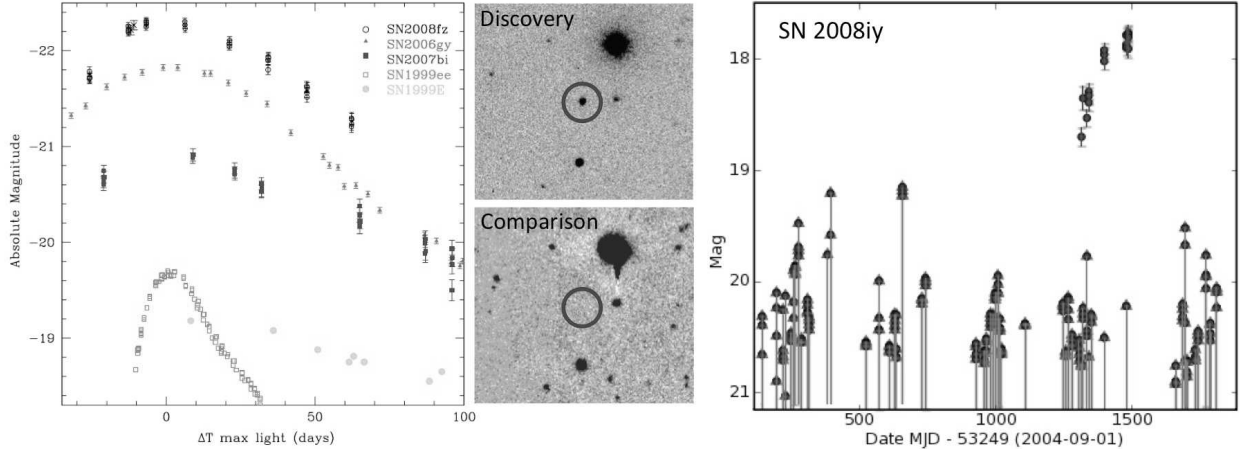}
\caption{ Examples of extreme SNe discovered by CRTS.  Left: The top points show
the light curve of SN 2008fz = CSS080922:231617+114248, the most luminous supernova
discovered until then (Drake et al. 2010).  Light curves of two other hyper-luminous
SNe, and two normal SNe are plotted below it.  The images show the discovery data
and the baseline comparison.  Note the absence of a visible host; it is a dwarf
galaxy with an absolute magnitude $M_R \approx -17$, comparable to that of the SMC.
Right:  Light curve of the extremely slow SN 2008iy = CSS080928:160837+041627.
This type IIn took $> 400$ days to reach the peak brightness, and occurred in an
extreme dwarf galaxy host with $M_R \approx -13$, i.e., less than $1/500^{th}$
of the Milky way.
}
\end{figure*}

\begin{figure*}[t]
\centering
\psbox[xsize=15.5cm]
{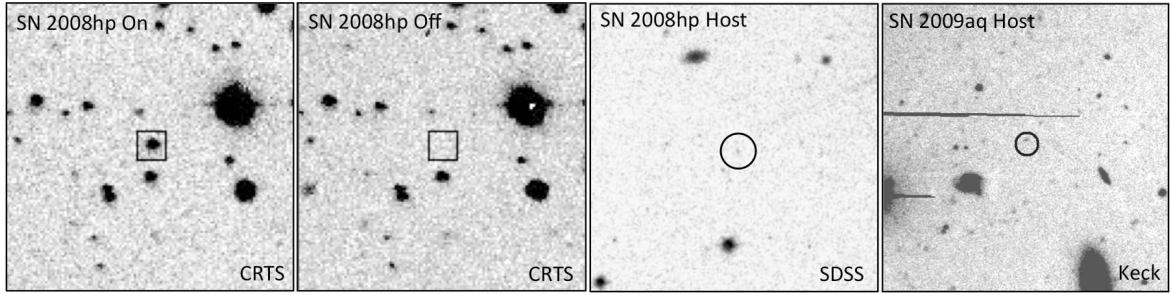}
\caption{ Further examples of the extreme dwarf galaxy hosts of luminous SNe.
The first two panels show the images of SN 2008hp = CSS081122:094326+251022
at the discovery epoch, and after it has faded away.  The next panel show a
zoom-in on the SDSS image of the field; the $\sim 23$ mag host galaxy is circled,
corresponding to the absolute magnitude $M_r \approx -12.7$.  The last panel shows
the confirmed $\sim 23$ mag host galaxy (circled) of SN 2009aq = CSS090213:030920+160505,
with the absolute magnitude $M_r \approx -13$. 
}
\end{figure*}

\begin{figure*}[t]
\centering
\psbox[xsize=16.5cm]
{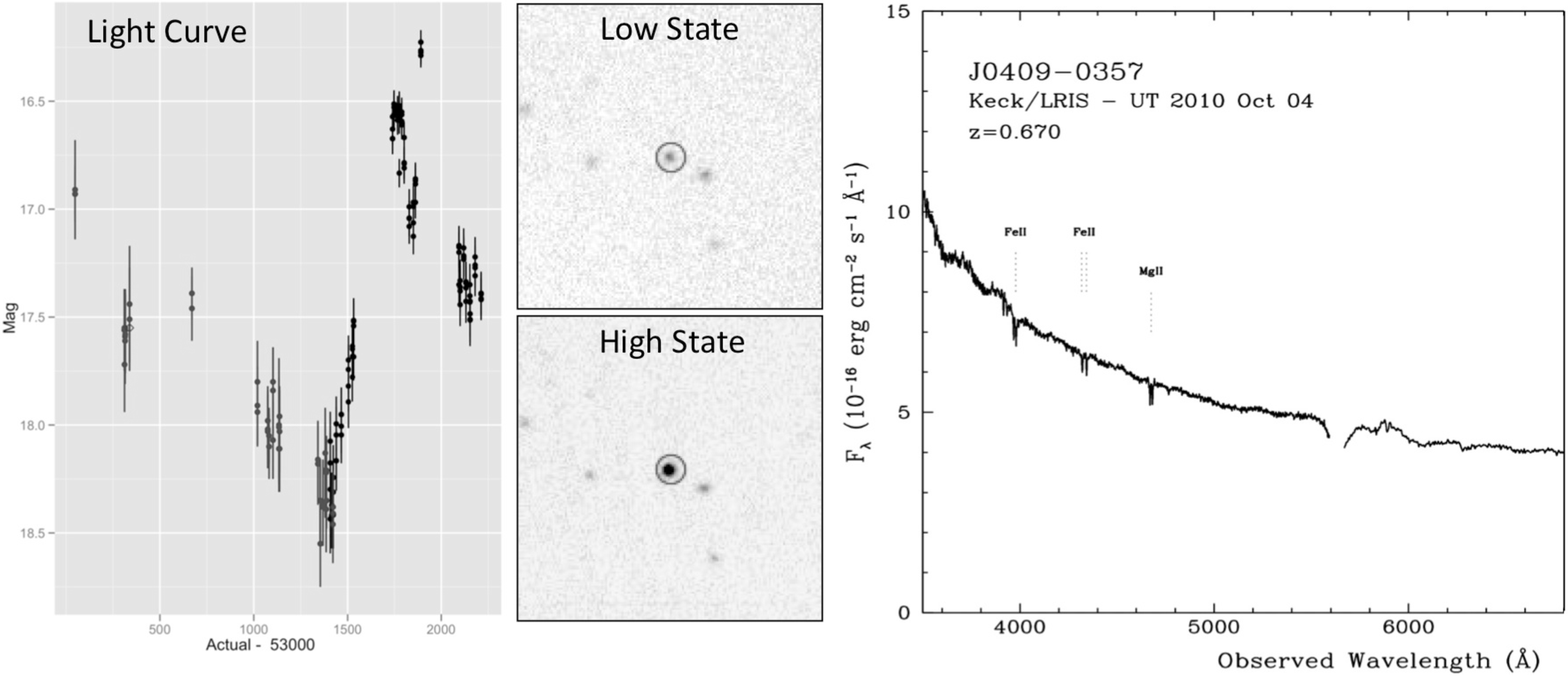}
\caption{ An example of a variability-based counterpart of a previously
unidentified $Fermi$ source 1FGL J0409-0357.  A composite light curve using the
data from both PQ and CRTS surveys, spanning almost 7 years, is shown on the left,
followed by the images illustrating the source in its low and high states.  This
is the most variable source in the error ellipse of the $Fermi$ source.  Its Keck
spectrum is shown on the right, with a featureless blue continuum, typical of
blazars, and absorption lines at $z = 0.670$, which may be due to the host galaxy
or to an unrelated absorber along the line of sight.  From Mahabal et al., in prep.
}
\end{figure*}

Our event triggering filter favors detection of SNe that greatly outshine their
host galaxies, since it is based on a flux contrast; this is the opposite
selection effect from many other SN surveys that require
a presence of an obvious host galaxy, and thus favor the more luminous hosts,
possibly biased against finding SNe from dwarf galaxies.  We already noted
(Drake et al. 2009) that we see substantial numbers of SNe from dwarf hosts,
often with an extremely low luminosity (Fig. 2).  The specific SN rates (per
unit stellar mass) in these systems may be orders of magnitude larger than
in the normal, $L_*$ galaxies (this finding was subsequently confirmed by
Neill et al. 2011).  Pending a more thorough analysis, there is a
hint that the overall SN rate may be underestimated.  An even more interesting
is a trend that hyper-luminous events seem to favor the low-mass hosts.  If 
this is borne out by a more detailed analysis, it would suggest a top-heavy
IMF in low-mass galaxies, which, in turn, may be caused by their lower
metallicities.  A similar effect was already noted for the GRB host galaxies
(see, e.g., Levesque et al. 2010, and refs.~terrain).  These local star-forming
dwarfs may be rough analogs to the first galaxies and their Pop.~III SNe.

Blazars, or, more generally, beamed AGN, offer numerous scientific opportunities:
they probe the physics of relativistic jets, AGN unification, they are perhaps
the dominant extragalactic $\gamma$-ray sources, and the dominant foreground 
radio source population for the modeling of CMBR fluctuations, etc.  Their
extreme variability offers a wavelength-unbiased mechanism for their discovery,
and, correlated between different wavelength regimes, it can be used to constrain
theoretical models.  CRTS is effectively conducting a statistical, almost-all-sky
monitoring campaign, an approach complementary to the standard targeted studies of
selected sources.  We are in the process of doing a correlative analysis between
CRTS, $Fermi$, and radio data from OVRO. To date, we have discovered several tens
of previously unknown blazars on the basis of their optical variability; this 
will provide a check on the selection effects affecting other discovery methods,
and lead to the more complete catalogs of these cosmic accelerators.
In particular, we are searching for the counterparts of unidentified $Fermi$
sources by obtaining archival light curves of all objects in their error ellipses,
covered by our CRTS and PQ images, with time baselines spanning $\sim 7$ years.
An example of such a variability-based ID is shown in Fig. 3.

\begin{figure*}[t]
\centering
\psbox[xsize=16.5cm]
{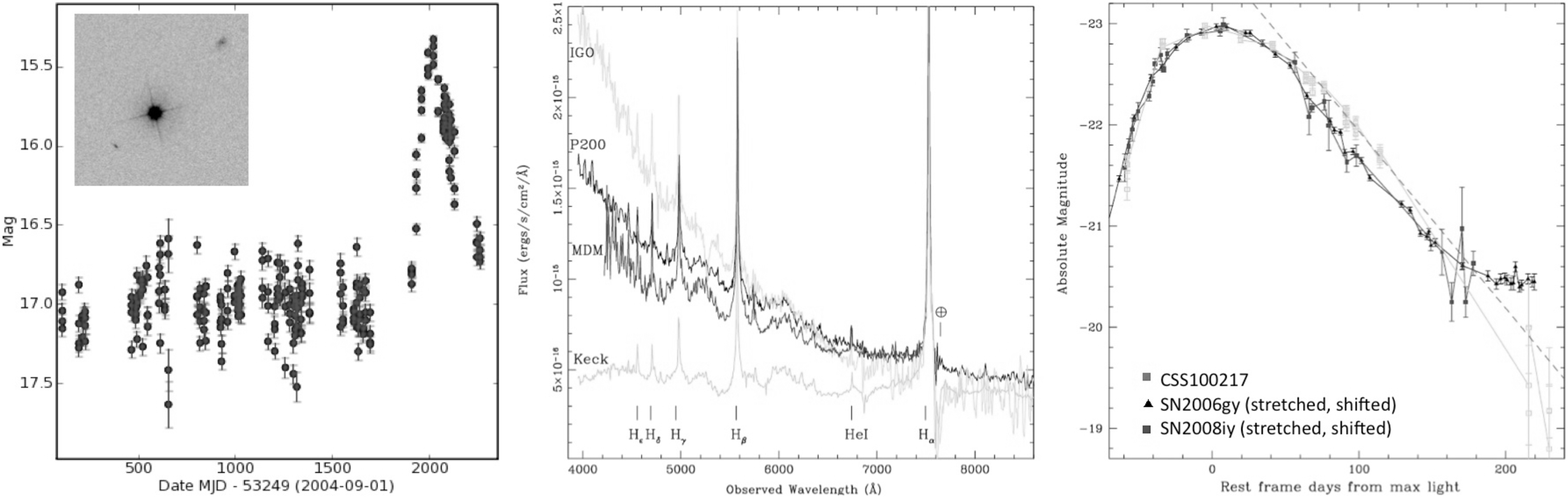}
\caption{ The unusual transient CSS100217:102913+404220, a possible hyper-luminous
Supernova from an AGN accretion disk, in a narrow-line Seyfert 1 galaxy at $z = 0.147$.
The CRTS light curve is on the left; the inset is the cutout from an HST image showing
a single, unresolved point source, indicating that the event occurred within $\sim 150$
pc from the AGN, i.e., well within the narrow-line region.  The middle panel shows a
sampling of the evolving spectra obtained during the spring of 2010; they are all
consistent with a combination of the pre-explosion AGN spectrum, and a type IIn SN.
The expanded light curve is shown on the right, compared with those of two other
hyper-luminous SNe IIn, scaled in brightness and with a time stretch applied.
From Drake et al. (2011b).
}
\end{figure*}

An intriguing transient CSS100217:102913+404220 was associated with a NLS1 
galaxy at $z = 0.147$ (Drake et al. 2011b; Fig. 4).  Its light curve is like that
of a SN IIn, but it would be the most luminous SN ever detected; the spectra are 
also consistent with a mix of the pre-explosion NLS1 AGN, and a SN IIn.
However, HST and Keck AO images do not resolve it from the AGN, implying that
the event occurred probably within $\sim 150$ pc of the nucleus, well within
the narrow-line region.  Ionizing radiation from the AGN should preclude any
star formation in its immediate vicinity, {\it except} in the shielded, outer,
cold regions of its accretion disk / obscuring torus (our data
exclude a significant extinction, making the disk origin more likely).
Massive star formation in the unstable outer parts of AGN accretion disks
has been long predicted (Shlosman \& Begelman 1987; see also Jiang \& Goodman 2011,
and refs.~therein).  This may be the first case of such a SN from an AGN
accretion disk, and more examples may be present in our archival data.

During three years of operation CRTS has discovered more than 500 dwarf nova type CVs,
thus increasing the total number of known systems by $\sim 25\%$.  The open data policy
of CRTS has led to numerous papers by other groups on these objects. Dedicated follow-up
of these CVs is now routinely undertaken within hours of discovery by a number of 
amateur and professional astronomers.  One significant discovery from this work is that
the activity cycle of such systems is linked to their intrinsic luminosity
(Wils et al.~2009). In Fig.~5 we show examples of three CVs discovered by CRTS.

\begin{figure*}[t]
\centering
\psbox[xsize=15.5cm]
{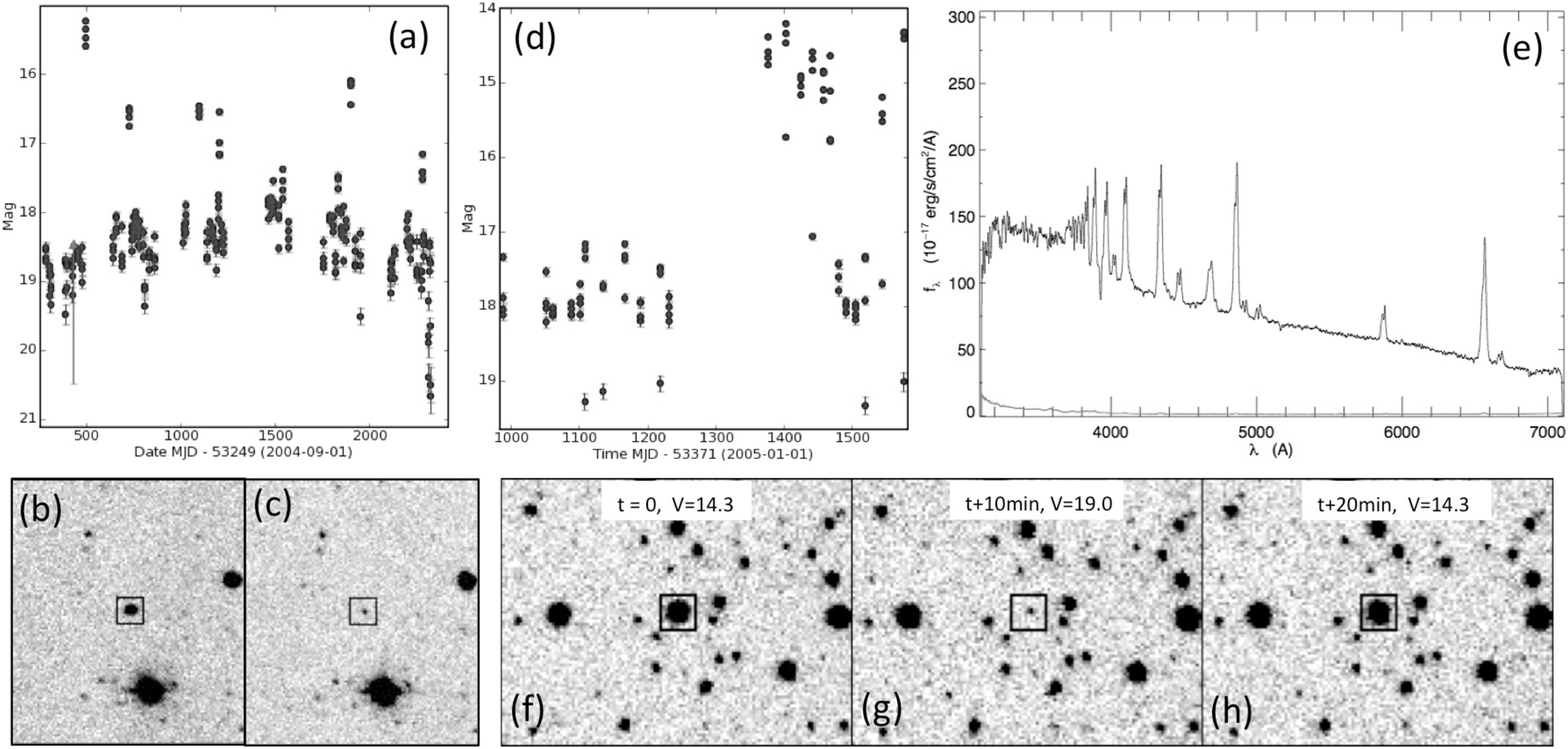}
\caption{Examples of CVs discovered by CRTS.
(a) The light curve of CSS091116:232551--014024, a typical CV found by our survey;
(b) and (c) show its images in high and low states.  (d) The light curve of a Polar
CV CSS081231:071126+440405; its spectrum, obtained at Palomar, is shown in (e),
displaying the typical strong H and He line emission.  (f,g,h) are consecutive
images, spaced by 10 min, of an eclipsing Polar CV CSS081231:071126+440405.
}
\end{figure*}

\begin{figure*}[t]
\centering
\psbox[xsize=15.5cm]
{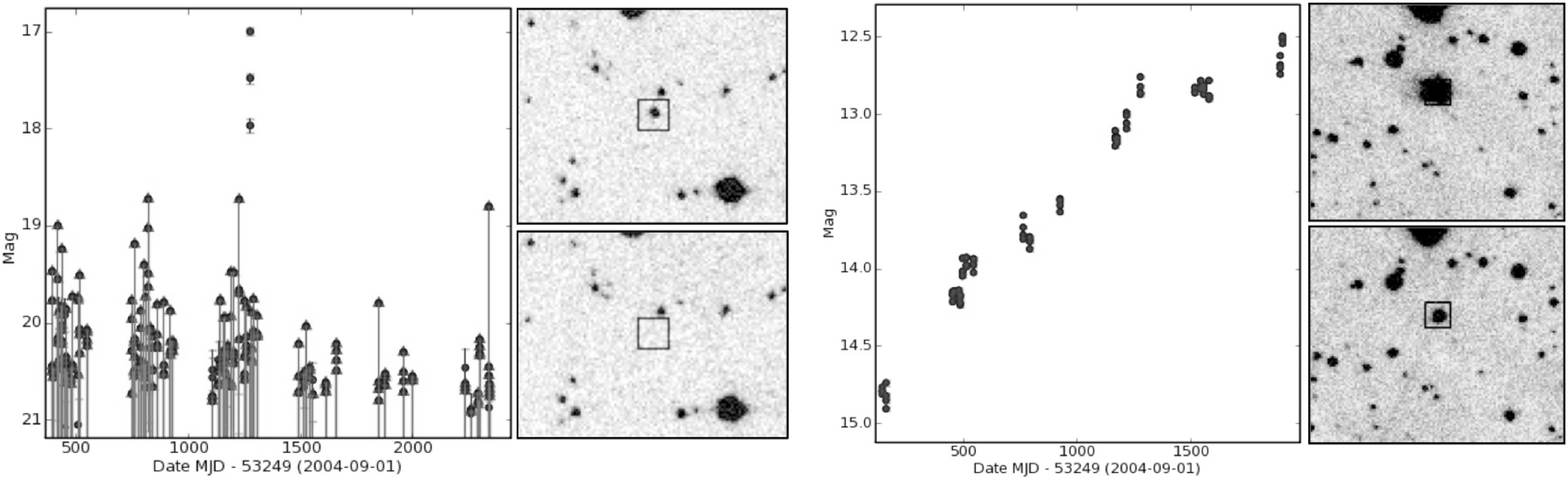}
\caption{Left:  An example of a flaring star light curve, CSS080228:044416+054730;
the adjacent images show the detection and the baseline.  Right:  A newly discovered
FU Ori object, CSS091110:060919Ð-064155 = IRAS 06068-Ð0641; the images on the right
show its brightening from 18 Jan 05 UT ($V \approx 14.8$ mag) to 10 Nov 09 UT
($V \approx 12.6$ mag).
}
\end{figure*}

The short (10 min) duration between the repeated observations in a sequence of CSS images
make the data ideal for discovery of high amplitude flares from UV Ceti stars. Such 
outbursts occur due magnetic reconnection in the atmospheres late dwarf stars and last 
for up to an hour. CRTS has discovered more than 100 flares, of which some outbursts 
rise more the 5 mag above the quiescence. The extreme variabilty of such systems is thought
to be a major contaminant to future transient surveys such as LSST as well as 
hampering efforts to discover planets around late type stellar systems. However, we
find that these flaring dwarfs are relatively easy to identify as such, using the
available archival data.  In Fig.~6
we present the light curve of one flare event that was observed directly before the flare
and three times as it declined.

The rapid cadence of CSS imaging has also led CRTS to the early serendipitous discovery
of eclipsing white dwarf systems (Drake et al.~2009).  In such cases, a secondary
companion as small as Earth can totally eclipse a white dwarf star. Faint secondaries
such as planets and brown dwarfs will naturally led to eclipses of many magnitudes
lasting less than $\sim 20$ minutes. Subsequent to initial eclipsing WD discoveries,
Drake et al.~(2011a) explored archival CSS data for more than ten thousand of white
dwarfs, and found more than a dozen eclipsing WD systems with low-mass companions.

In addition, we see a number of variable stars of various types, including young and
extremely rare FU Ori type stars, and high proper motion stars (since their new positions
differ from the baseline comparison images.

While we do have an active follow-up program at Palomar, Keck, at various telescopes in
Chile and elsewhere, and we have developed a broad, international network of
collaborations to this end.  However, the scientific output of CRTS is currently limited
by the lack of the follow-up, with only a small fraction of the transients covered.
This bottleneck (especially in spectroscopy) can only get worse, as more and larger
synoptic surveys come on line.

This brief account is just indicative of the wealth of data produced by CRTS and the
possible resulting projects.  Our open-data policy benefits the entire astronomical
community, generating science now, and preparing us for the larger surveys to come.

\section{Automated Classification of Transients}

Perhaps the key technological challenge facing synoptic sky surveys is the
automated classification of transients, which then feeds their follow-up
prioritization.  This is a very challenging problem, which we have been
tackling for some years now (see, e.g., Donalek et al. 2008, Mahabal et al.
2008ab).  We are currently exploring a number of possible methods, using the
CRTS data stream as a testbed.  In particular, we are deploying a novel
``citizen science'' project, $SkyDiscovery.org$, whose aim is to harvest the
human pattern recognition skills and domain expertise, and turn it into
scalable algorithms for automated rejection of artifacts and classification
of genuine transients.  A more detailed account of these efforts is outside
of the scope of the present paper, and will be reported elsewhere.

\vspace{1pc}
\noindent 
We wish to thank numerous collaborators who have contributed to the survey and its
scientific exploitation so far.
CRTS is supported by the NSF grant AST-0909182, and in part by the Ajax
Foundation.  The initial support was provided by the NSF grant AST-0407448, and some
of the software technology development by the NASA grant 08-AISR08-0085. 
The analysis of the blazar data was supported in part by the NASA grant 08-FERMI08-0025. 
Education and public outreach activities are supported in part by the Microsoft Research
WorldWide Telescope team.  The CSS survey is supported by the NASA grant NNG05GF22G.
We are grateful to the staff of Palomar, Keck, and other pertinent observatories for
their expert help during our follow-up observations.  Event publishing and analysis
benefits from the tools and services developed by the U.S. National Virtual Observatory
(now Virtual Astronomical Observatory).  SGD wishes to than the workshop organizers
for the invitation and hospitality during the workshop.

\section*{References}

\re
Djorgovski, S.G., Brunner, R.J., Mahabal, A.A., Odewahn, S.C., de Carvalho,
R.R., Gal, R.R., Stolorz, P., Granat, R., Curkendall, D., Jacob, J., \&
Castro, S. 2001, in {\sl Mining the Sky}, eds. A.J. Banday et al.,
ESO Astrophys. Symp., p. 305, Berlin: Springer Verlag
\re 
Djorgovski, S.G., Baltay, C., Mahabal, A., Drake, A.J., Williams, R.,
Rabinowitz, D., Graham, M., Donalek, C., Glikman, E., Bauer, A., Scalzo, R.,
Ellman, N., \& Jerke, J. 2008, AN, 329, 263
\re
Donalek, C., Mahabal, A., Djorgovski, S.G., Marney, S., Drake, A.J., Graham, M.,
Glikman, E., \& Williams, R. 2008, in {\sl Proc. Intl. Conf. on Classification
and Discovery in Large Astronomical Surveys}, AIP Conf. Ser., 1082, 252
\re
Drake, A.J., Djorgovski, S.G., Mahabal, A., Beshore, E., Larson, S., Graham, M.,
Williams, R., Christensen, E., Catelan, M., Boattini, A., Gibbs, A., Hill, R., 
\& Kowalski, R. 2009, ApJ, 696, 870
\re
Drake, A.J., Djorgovski, S.G., Prieto, J., Mahabal, A., Balam, D., Williams, R.,
Graham, M., Catelan, M., Beshore, E., \&  Larson, S. 2010, ApJ, 718, L127
\re
Drake, A.J., Beshore, E., Catelan, M., Djorgovski, S.G., Graham, M., Kleinman, S.,
Larson, S., Mahabal, A., \& Williams, R. 2011a, ApJ, in press 
\re
Drake, A.J., Djorgovski, S.G., Mahabal, A., et al. 2011b, ApJ, in press 
\re
Gal-Yam, A., et al. 2009, Nat, 462, 624
\re
Harwit, M. 1981, Cosmic Discovery, New York: Basic Books
\re
Jiang, Y., \& Goodman, J. 2011, ApJ, submitted (arXiv/1011.3541)
\re
Levesque, E., Kewley, L., Berger, E., \& Zahid, H. 2010, AJ, 140, 1557
\re
Mahabal, A., Djorgovski, S.G., Turmon, M., Jewell, J., Williams, R., Drake, A.J., 
Graham, M., Donalek, C., Glikman, E., and the PQ Team 2008a, AN, 329, 288
\re
Mahabal, A., Djorgovski, S.G., et al. 2008b, in
{\sl Proc. Intl. Conf. on Classification and Discovery in Large Astronomical
Surveys}, AIP Conf. Ser., 1082, 287
\re
Neill, J.D., et al. 2011, ApJ, in press (arXiv/1011.3512)
\re
Paczynski, B. 2000, PASP, 112, 1281
\re
Shlosman, I., \& Begelman, M. 1987, Nat, 329, 810
\re
Williams, R., Djorgovski, S.G., Drake, A., Graham, M., \& Mahabal, A. 2009,
in {\sl Proc. ADASS XVII}, eds. J. Lewis et al., ASPCS, 411, 115
\re 
Wils, P., Gaensicke, B. T., Drake, A.J., \& Southworth, J. 2010, MNRAS, 402, 436

\label{last}

\bigskip
\noindent\rule[2mm]{70mm}{0.2mm}

\noindent
This review will appear in proc.
{\sl The First Year of MAXI: Monitoring Variable
X-ray Sources}, eds. T.~Mihara \& N.~Kawai, Tokyo: JAXA Special Publ. (2011).

\end{document}